# Magneto-radiative modelling and artificial neural network optimization of biofluid flow in a stenosed arterial domain

S P Shivakumar[1], Gunisetty Ramasekhar[2], P Nimmy[1], S Areekara[1], L Thanuja[3], T V Smitha[1], S Devanathan[4], Ganesh R Naik[5], K V Nagaraja[1]

[1]Computational Science Lab, Amrita School of Engineering, Amrita Vishwa Vidyapeetham, Bengaluru, 560035, India

[2]Department of Mathematics, Rajeev Gandhi Memorial College of Engineering and Technology (Autonomous), Nandyal, 518501, Andhra Pradesh, India

[3]Department of Student Affairs, Amrita Vishwa Vidyapeetham, Bengaluru, 560035, India

[4]Department of Mechanical Engineering, Amrita School of Engineering, Amrita Vishwa Vidyapeetham, Bengaluru, 560035, India

[5]College of Medicine and Public Health, Flinders University, Adelaide, SA 5042, Australia

[*]**Corresponding Author's Email**: kv_nagaraja@blr.amrita.edu

**Abstract:** The increasing complexity of cardiovascular diseases and limitations in traditional healing methods mandate the invention of new drug delivery systems that assure targeted, effective, and regulated treatments, contributing directly to UN SDGs 3 and 9, thereby encouraging the utilization of sustainable medical technologies in healthcare. This study investigates the flow of a Casson-Maxwell nanofluid through a stenosed arterial domain. The quantities, such as skin friction and heat transfer rate, are analysed in detail. The Casson-Maxwell fluid shows a lower velocity profile than the Casson fluids, which indicates the improved residence time for efficient drug delivery. The heat transfer rate shows an increase with higher volume fractions of copper and aluminium oxide nanoparticles and a decrease with higher volume fractions of silver nanoparticles. The skin friction coefficient decreases by 219% with a unit increase in the Maxwell parameter, whereas it increases by 66.1% with a unit rise in the Casson parameter. This work supports SDGs 4 and 17 by fostering interdisciplinary learning and collaboration in fluid dynamics and healthcare innovation. Additionally, the rate of heat flow was forecasted (with an overall R-value of 0.99457) using the Levenberg-Marquardt backpropagation training scheme under the influence of magneto-radiative, linear heat source and Casson-Maxwell parameters along with the tri-metallic nanoparticle volume fractions. It is also observed that the drag coefficient is most sensitive to the changes in the Maxwell parameter.

# 1. Introduction

Blood is an essential biofluid that circulates through arteries and veins to eliminate waste products from cells and supply nutrients and oxygen to the tissues and organs. Arterial blood flow has emerged as a significant area of research for medical professionals and engineers, owing to its relevance in the management of cardiovascular disorders and the innovation of cardiac devices. Accumulation of fatty materials, cholesterol, and cellular waste products in the arteries reduces the lumen of the arteries, affecting the blood flow, a condition known as stenosis. This increases the flow resistance, elevating the blood pressure and reducing oxygen supply to numerous parts of the body, leading to the development of major health issues such as peripheral artery disease, stroke, and heart attack. Thus, to avoid major cardiac problems, stenosed arteries must be diagnosed and treated without any delay. This makes a substantial contribution to SDG3, which seeks to ensure everyone's well-being and health.

Blood is often regarded as a non-Newtonian fluid because of its shear-thinning properties. The shear-thinning fluid models widely applied in biomedical applications include the non-Newtonian fluid Casson and Maxwell models (see [1]). Since a single model would not be able to anticipate every characteristic of non-Newtonian materials, researchers have offered several non-Newtonian models. Integral, rate, and differential type fluid models are the three main categories into which these models can be divided. The fluid model studied here belongs to the class of rate-type fluids known as Maxwell fluids (M-F). Maxwell introduced the concept of the Maxwell fluid. Because of its distinct behaviour in reaction to applied forces, the Maxwell fluid's rheology differs from other non-linear models. The viscoelastic features of the Maxwell fluid, in contrast to other models, allow it to behave as both a liquid and a solid under certain circumstances.

The Casson fluid (C-F) model, a non-Newtonian fluid with yield stress, is frequently used to simulate blood flow in small arteries. Several researchers have mathematically modelled blood flow in narrow arteries at low shear rates using the C-F model. The heat transmission of Casson hybrid nanofluid passing through the constricted artery was examined by Ezhilarasi and Mohanavel [2]. The role of nanoparticles on Casson nanofluid blood flow via the cardiac artery was explored by Ramasekhar et al. [3]. Ausaru et al. [4] deliberated the solute dispersion and C-F flow over catheterized stenosed arteries. The solute dispersion in Casson blood circulation through a stenosed artery with a stiff permeable wall was elucidated by Dompok and Jaafar [5]. To investigate the heat transmission over a Riga plate, Khan et al. [6] considered the two-dimensional flow of a Maxwell fluid. They found that the momentum layer thickness dropped as the Deborah number increased in terms of the Maxwell parameter. Together with the consequences of viscous dissipation, thermal radiation, and magnetic field, the dynamics of Maxwell nanofluid movement over an elongating cylinder were investigated by Faraz and Park [7]. Their study demonstrated that as the Maxwell fluid parameter escalates, the horizontal fluid velocity also increases. According to research by Din et al. [8] on the mass and heat transfer of M-F with improved variable thermal characteristics across a porous sheet, they proved that M-F enhances heat transfer efficiency and keeps temperatures steady under high-temperature drilling settings. In a Darcy porous medium, Alrihieli et al. [9] examined the combined influence of many essential factors on the hydrodynamic behavior of M-F over a Riga plate situated. They concluded that heat and mass transmission rates are escalated by larger values of the Maxwell parameter. In their investigation of the MHD channelized M-F flow, Ahmad et al. [10] considered the simultaneous effects of chemical reactions, heat production, and suction/injection. Their study showed that when the Maxwell parameter amplified, the velocity decreased, indicating that viscoelastic liquids with longer relaxation durations had less flow flexibility.

The combined impacts of viscoelastic behaviour and non-Newtonian viscosity, Casson-Maxwell fluid flow, accelerating mass transfer rates. This is especially helpful in procedures like reactive distillation systems and multiphase reactions, where mass transfer constraints are important. Moreover, using Casson-Maxwell fluid models to deliver drugs better increases the efficacy and efficiency of pharmacological treatments that target the tissues. Assiri et al. [11] analysed the MHD Casson-Maxwell nanofluid movement along the gap between a cone and a disk. Over a stretching surface, the flow at the stagnation point of Casson-Maxwell fluid was studied by Shah et al. [12]. Rao et al. [13] examined the effects of nanoparticles on the heat and mass transport of upper-convected Casson and Maxwell fluid above a stretched sheet. Across a porous stretched sheet, Jayaprakash et al. [14] examined how several non-dimensional characteristics affected the non-Newtonian Casson-Maxwell fluid's steady flow. Islam et al. [15] explored the dynamic properties of periodic magnetohydrodynamic C-F and M-F, focusing on the interaction of mass transfer and heat processes under the impact of non-linear radiation and Arrhenius activation energy.

With major advancements in tissue engineering, targeted drug delivery, therapeutic interventions, and diagnostics, nanomedicine is a young discipline that capitalizes on the special qualities of nanomaterials to transform healthcare. Nanoparticles in a basic liquid make up the colloidal suspensions called nanofluids. Nanofluids have potential applications in the medical domain, the petroleum industry, car radiators, heat pipes, solar collectors, and heat exchangers. Choi et al. [16] found that adding a relatively small number of nanotubes to a base fluid results a notable rise in its effective thermal conductivity; at around 1 vol% nanotubes, the thermal conductivity ratio surpasses 2.5. Also, researchers have developed blood flow mathematically, and their most recent innovation employs nanoparticles in blood for effective drug delivery. Along a narrowed segment of a stenosed artery, Dolui et al. [17] investigated the flow of hybrid nanofluid. They discovered that interfacial nanolayers significantly alter blood temperature in order to prevent and treat heart disease. Algehyne et al. [18] concentrated on employing blood-hybrid nanofluid flow with gold-tantalum nanoparticles in a tilted cylindrical artery with composite stenosis. Berkan et al. [19] explored heat transfer and nanofluid peristaltic flow in drug delivery devices utilizing analytical and numerical method. Waqas et al. [20] used CFD to examine the flow behaviour of a hybrid nanofluid using blood as the base fluid over a stenosed artery. They discovered that silver and gold nanoparticles work well as a medication to lower the hemodynamics of stenosis. Arif et al. [21] studied the heat flow analysis of the couple stress Casson trihybrid nanofluid using dissimilarly shaped nanoparticles in blood for biomedical purposes. Asha and Srivastava [20] explored the blood flow in the presence of various shapes of nanoparticles through curved arteries and concluded that spherical nanoparticles facilitate better thermal performance.

The study of the intricate relationship between the magnetic field and electrically conducting fluids is known as magnetohydrodynamics (MHD). Blood velocity tends to decrease when a magnetic field is enforced in the perpendicular direction of blood flow because of the Lorentz force. The use of this principle in magnetic resonance imaging (MRI) produces finely detailed pictures of internal structures, which are highly advantageous for diagnosis and treatment planning. The advancements in the production of targeted drugs and MRI lead to increased industrial capacity and improved healthcare facilities, both of which support SDG 9. Alraddadi et al. [21] assessed the behavior of blood flow through a stenosed artery when exposed to an angled magnetic field. They concluded that since in the blood the charged particles undergo the Lorentz force, which slows down their motion, stronger magnetic fields may result in lower blood flow rates. Using a Casson fluid model with gold nanoparticles and considering MHD effects, Azmi et al. [22] utilized the fractional derivative techniques to study blood nanofluid flow in a slip cylinder. Motivated to understand the causes underlying cardiovascular illness better and advance medicinal applications, through an inflamed stenosed artery, Turabi et al. [23] examined the heat transfer and flow properties of MHD blood hybrid nanofluid flow. Noranuar et al. [24] investigated how the unsteady MHD flow in a revolving channel across a porous media was affected by suspending carbon nanotubes in human blood, which was modeled as a Casson nanofluid. Nazar and Shabbir [25] investigated the electromagnetohydrodynamic movement of Maxwell fluid along a stenosed artery.

Thermal radiation refers to the electromagnetic waves emitted by a substance due to its temperature. The nature of this radiation is directly influenced by the temperature of the transmitting material [26]. The influence of the radiation process intensifies as the temperature differential between the liquid's surface and surroundings rises. The heat transport rate will ultimately be impacted by this change as well. By the 1970s, the physiological foundation for the advantages of hyperthermia with contemporary cancer treatments, such as chemotherapy and radiation, had begun to be recognized. Hyperthermia was identified as the perfect supplemental treatment for radiation therapy. In an inclined stenotic artery, Kefene and Rikitu [27] investigated the hematocrit-dependent flow of blood nanofluid while considering heat radiation and magnetic field effects. The role of chemical reactions and heat radiation on the MHD of a blood based nanofluid via a stenosed artery was examined by Imoro et al. [28]. The role of heat radiation on the blood hybrid nanofluid flow in a tilted cylindrical W-shaped symmetric stenosis artery was investigated by Algehyne et al. [29]. Omama et al. [30] explored how the flow of Sisko nanofluid via constricted arteries was affected by a novel Tetra-HNF model in addition to thermal radiation effects. Using Rabinowitsch fluid model, Khaliq et al. [31] investigated how heat radiation affected the blood's shear-thinning properties in nonuniform inclined permeable stenosed arteries.

 Artificial Neural Networks, or ANNs, have emerged as a crucial artificial intelligence method in recent years. By mimicking the interconnected activity of brain cells, neural network technology can learn, identify patterns, make predictions, and solve problems across numerous domains. ANNs are becoming increasingly popular because of their remarkable capacity to solve challenging and highly non-linear mathematical problems. In complicated domains, including fluid dynamics, biotechnology, and biological computing, ANNs offer a versatile computational framework that is very helpful. Across a curved stretched surface, Ul-Haq et al. [32] used numerical simulation using an ANN technique to forecast the non similar solution of MHD transport of nanofluid. Using experimental data, Topal et al. [33] demonstrated how to use ANN and Genetic Algorithm to forecast the viscosity values of the nanofluid at various temperatures. In Khan et al.'s [34] study, ANNs are created to predict the flow of nanofluids with numerous slips. Shafiq et al. [35] have taken into consideration MHD micropolar nanofluid flow towards the stretched surface using an ANN. Using ANN, Paul et al. [36] examined ballooned catheterization along a bloodstream carrying nanoparticles over a stenotic arterial cavity.

This work investigates the magneto-radiative behaviour of Casson-Maxwell nanofluid flow through stenosed artery modelled as a stretching sheet, incorporating a thermal radiation, magnetic field, a linear heat source, and metallic nanoparticles. The study focuses on understanding how flow resistance, heat transfer, and wall shear stress influence the efficiency of targeted drug delivery and medication in constricted biological domains. The findings contribute to the development of precise, non-invasive therapeutic systems and align with global healthcare innovation goals outlined in the UN SDGs, particularly promoting good health, innovation, and sustainable infrastructure. This study seeks to explore the following key research questions:

- How do Casson-Maxwell rheological properties influence the velocity of nanofluid flow in stenosed arteries?
- What parameter combinations minimize harmful wall shear stress while ensuring drug deposition?
- What are the optimal conditions for hyperthermia-assisted drug delivery?

- How can the transport phenomenon be trained using artificial neural networks to optimize targeted drug release and medication?

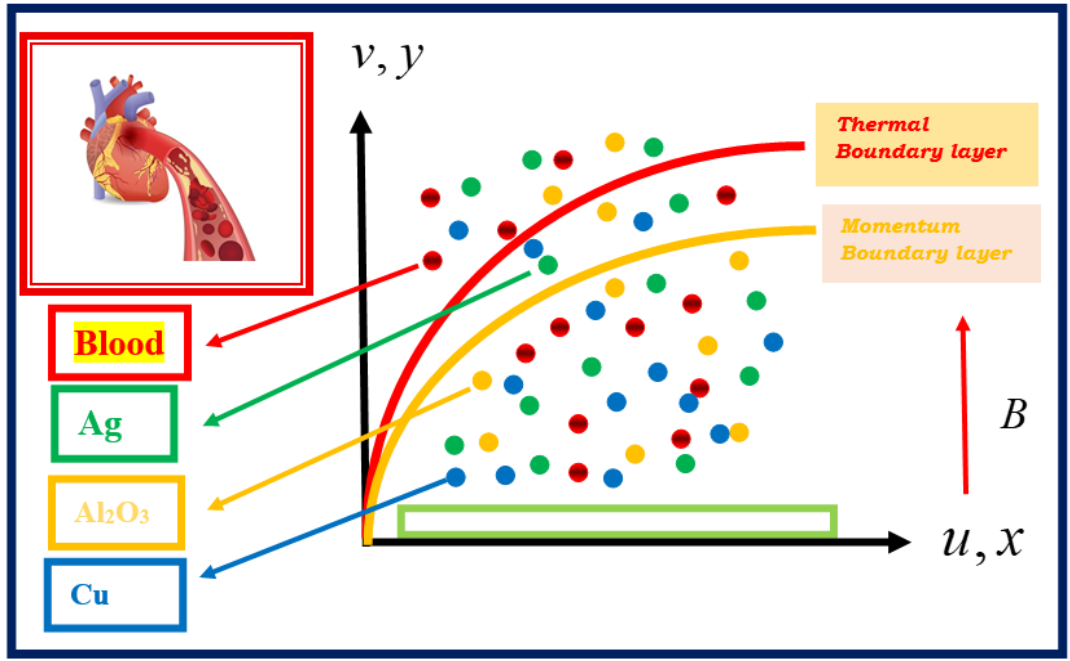

**Fig. 1.** Flow geometry of the model.

## 2. Mathematical model

The 2-dimensional steady flow of a C-M NF over a stretched surface is studied, wherein the stenosed artery is idealized and modeled as a stretching sheet. The $u$ and $v$ represent velocity components of the fluid motion in the $x$ and $y$-axis, the sheet's velocity is given as $U = ax$, and $\lambda^*$ is the time relaxation constant. The sheet's temperature and the free stream are denoted by $T_w$ and $T_\infty$, respectively.

The rheological formula for an incompressible movement through a Casson-type fluid follows [37,38]:

$$\tau_{ij}^* = \begin{cases} 2\left(\mu_b^* + \dfrac{p_y^*}{\sqrt{2\pi^*}}\right)e_{ij}^*, & \pi^* > \pi_c^* \\ 2\left(2\mu_b^* + \dfrac{p_y^*}{\sqrt{2\pi_c^*}}\right)e_{ij}^*, & \pi^* \leq \pi_c^* \end{cases}$$

Here $\pi^* = (e_{ij}, e_{ij})$ product of element of the deformation rate, $\pi_c^*$ represents a critical value of this item according to the Casson fluid model, $p_y^*$ represents a yield stress, $\mu_b^*$ represents a plastic dynamic viscosity of the Casson fluid model. The above equation takes the following form when $\pi^* \leq \pi_c^*$: $\tau_{ij}^* = 2\mu_b^*\left(1 + \dfrac{1}{\beta}\right)e_{ij}^*$, where the Casson fluid parameter is $\beta = \dfrac{\mu_b^* \sqrt{2\pi_c^*}}{p_y^*}$.

The stress tensor $\tau_{ij}$ for the Maxwell fluid can be associated with the deformation rate tensor $d_{ij}$, as [39]

$$\lambda \frac{\Delta \tau_{ij}}{\Delta t} + \tau_{ij} = 2\eta d_{ij}$$

where λ the relaxation time and η the viscosity coefficient. The time derivative Δ/Δt used in the above equation refers to the upper convected time derivative, which is formulated to satisfy the requirement of continuum mechanics (i.e., frame indifference and material objectivity). When this time derivative is applied to the stress tensor, it takes the following form [39]:

$$\frac{\Delta \tau_{ij}}{\Delta t} = \frac{D\tau_{ij}}{Dt} - L_{jk}\tau_{ik} - L_{ik}\tau_{kj}$$

where $L_{ij}$ is the velocity gradient tensor.

Thermophysical properties of Ternary hybrid nanofluids are

$$Z_1 = \frac{\mu_{Thnf}}{\mu_f}, Z_2 = \frac{\rho_{Thnf}}{\rho_f}, Z_3 = \frac{(\rho c_p)_{Thnf}}{(\rho c_p)_f}, Z_4 = \frac{k_{Thnf}}{k_f}, Z_5 = \frac{\sigma_{Thnf}}{\sigma_f}.$$

Dimensional form of $C_f$, and $Nu$ are defined by

$$C_f = \frac{\tau_w}{\rho_f U^2} \qquad (13)$$

$$Nu = \frac{x q_w}{k_f (T_w - T_\infty)} \qquad (14)$$

## 3. Computational Methodology and Validation

The non-linear boundary value problem governing the flow of a C-M NF through a stenosed artery, under the impact of the linear heat source, thermal radiation, and a magnetic field, is solved numerically using MATLAB software. The governing PDEs are transformed into coupled, non-linear ODEs using proper similarity transformations. These ODEs are then solved utilizing the bvp4-c solver in MATLAB, which is suited for boundary value problems with complex and stiff systems. An initial guess satisfying the boundary conditions is given, and mesh refinement is done adaptively to ensure convergence. Furthermore, Table 2 validates the accuracy of numerical scheme employed in this study. Let

$$y_1 = F, y_2 = F', y_3 = F'', y_4 = \theta, y_5 = \theta'$$

**Table 1: Ternary nanoparticles properties** *[43,44]*.

| Property | Blood | $Cu$ | $Ag$ | $Al_2O_3$ |
|---|---|---|---|---|
| Density $\rho$ ($kgm^{-3}$) | 1050 | 8933 | 10,500 | 3970 |
| Specific heat $C_p$ ($Jkg^{-1}K^{-1}$) | 3617 | 385 | 235 | 765 |
| Electrical conductivity $\sigma$ $(\Omega m)^{-1}$ | 0.8 | $5.96 \times 10^7$ | $6.3 \times 10^7$ | $35 \times 10^6$ |
| Heat conductivity $k_f$ ($Wm^{-1}K^{-1}$) | 0.52 | 401 | 429 | 40 |
| $Pr$ | 21 | - | - | - |

**Table 2:** Comparison table of $Nu_x(Re_x)^{-1/2}$ of the current study for various $Pr$ when all the parameters are zero.

| $Pr$ | $Nu_x(Re_x)^{-1/2}$ | | |
|---|---|---|---|
| | Ramzan et al., [41] | Yahya et al., [45] | Present result |
| 0.7 | 0.4560 | – | 0.4544472 |
| 2.0 | 0.9113 | 0.9112 | 0.9113576 |
| 6.13 | – | 1.7597 | 1.8954005 |
| 20.0 | – | 3.3540 | 3.3539083 |

## 4. Parametric analysis

This study investigates the hemodynamic and thermal characteristics of a C-M NF flowing through a stenosed artery modeled as a stretching sheet under the influence of thermal radiation, linear heat source, and magnetic field. The blood-based nanofluid is enhanced with $Ag, Cu,$ and $Al_2O_3$ nanoparticles to simulate targeted therapeutic and diagnostic applications. The governing boundary layer equations, numerically solved using MATLAB's bvp4c solver, reveal the significant effect of different physical parameters on temperature, velocity, wall shear stress profiles, and heat transfer rate. These findings are illustrated through graphs, demonstrating the influence of dimensionless parameters on physiological transport phenomena. The results provide valuable insights for optimizing hyperthermia-based stenosis treatment, magnetically guided drug delivery, and thermal management in vascular devices, thereby supporting the broader goal of improving biomedical intervention strategies in stenosed arteries.

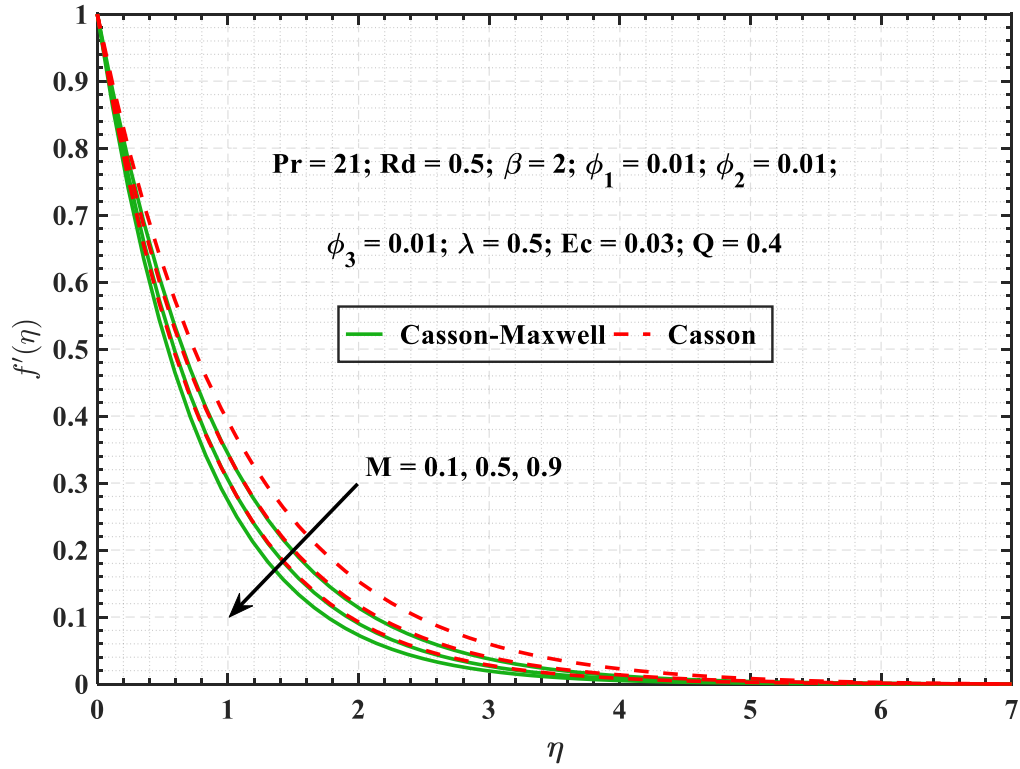

**Fig. 2.** $f'(\eta)$ versus $M$.

The graph (Fig. 2) compares the velocity profiles $F'(\eta)$ of Casson and C-M NF under varying magnetic parameter ($M$). It is observed that C-M NF exhibits a lower $F'(\eta)$ than the Casson fluid. This reduction in $F'(\eta)$ for the C-M NF is due to the combined effects of viscoelasticity (Maxwell model) and non-Newtonian behavior (Casson model). The Maxwell parameter introduces an elastic response that resists rapid changes in fluid motion, while the Casson parameter accounts for yield stress, both of which serve to curtail the flow. Since the C-M NF has a lower $F'(\eta)$ than the Casson fluid, this study adopts the C-M NF model to optimize targeted drug delivery applications.

The graph (Fig. 2) reveals that the increase of $M$ causes a fall in $F'(\eta)$ of the C-M NF. This is because of Lorentz force that appears owing to magnetic field, which gives the fluid a hydraulic resistance. The graph (Fig. 3) explains that when the Maxwell parameter ($\lambda$) increases, the $F'(\eta)$ of the C-M NF decreases. This is due to the increase in the viscoelastic nature of fluid, which results in the fact that the fluid is resistant to rapid deformation. An increase in Casson parameter ($\beta$), the $F'(\eta)$ of the C-M NF decreases throughout the flow domain. This happens because a greater $\beta$ characterizes a more explicit yield stress, which shows the blood's non-Newtonian behavior. A larger yield stress leads to a more significant fluid resistance to motion, and thus, the $F'(\eta)$ decreases. In targeted drug delivery, especially in stenosed arteries, slower flow enhances nanoparticle residence time, improving drug absorption and controlled release. In this way, one can regulate the $M, \lambda$, and $\beta$ of nanofluid to achieve efficient delivery through stenosed arteries by creating a more effective interaction between nanoparticles and stenosis.

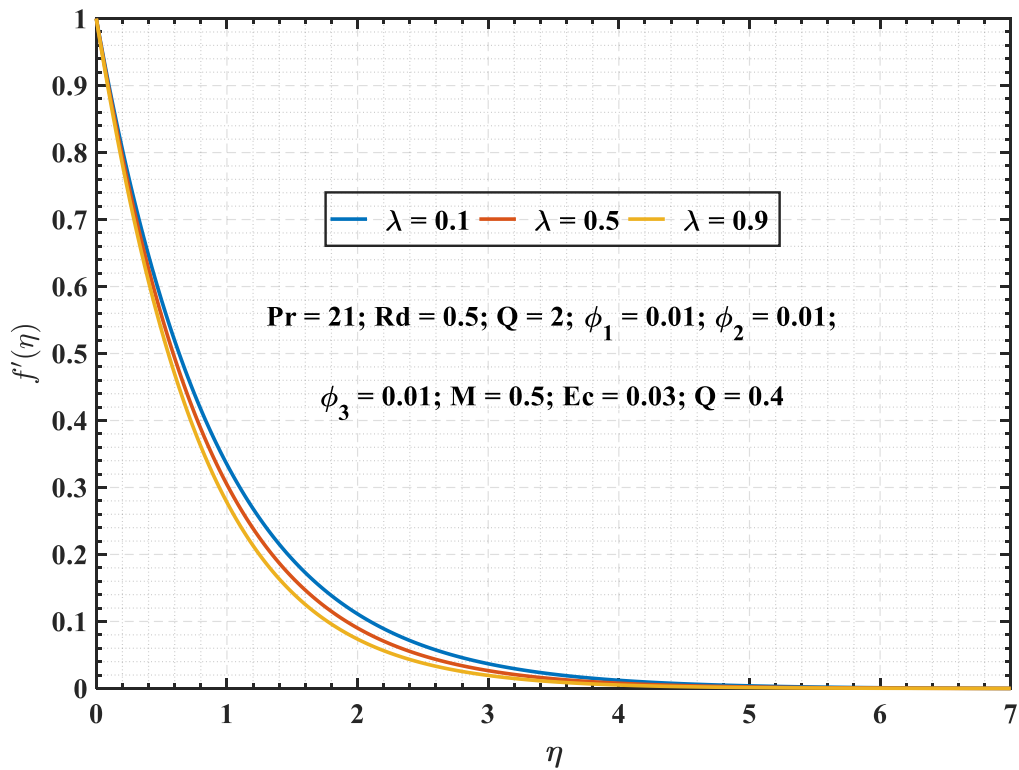

**Fig. 3.** $f'(\eta)$ versus $\lambda$.

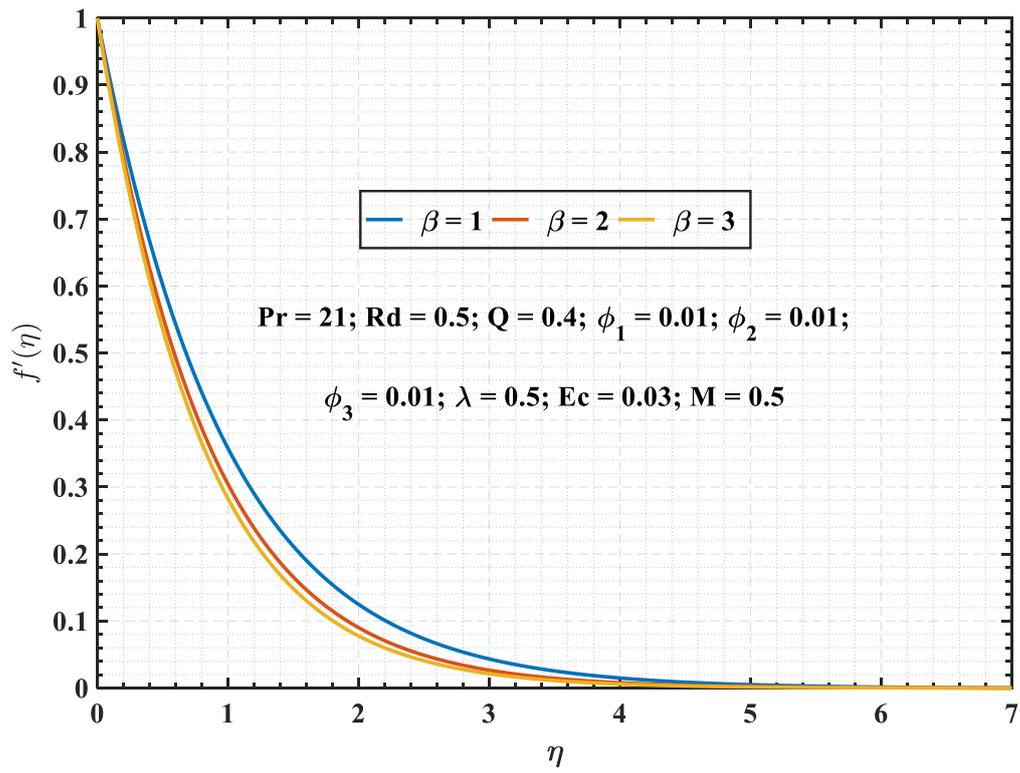

**Fig. 4.** $f'(\eta)$ versus $\beta$.

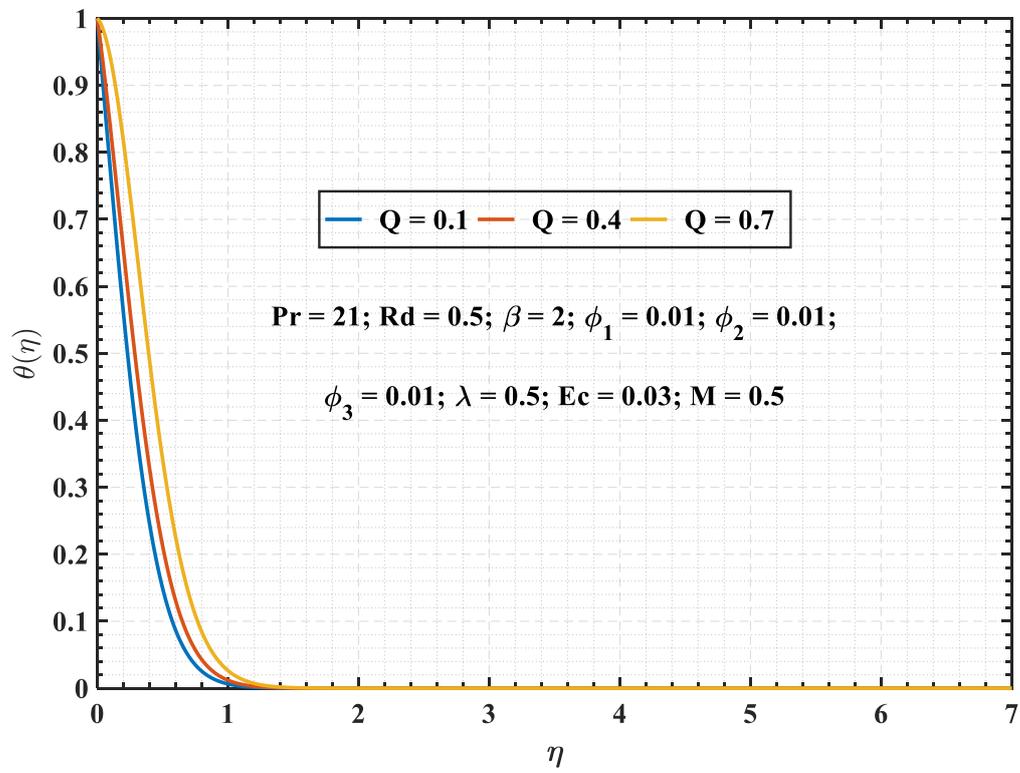

**Fig. 5.** $\theta(\eta)$ versus $Q$.

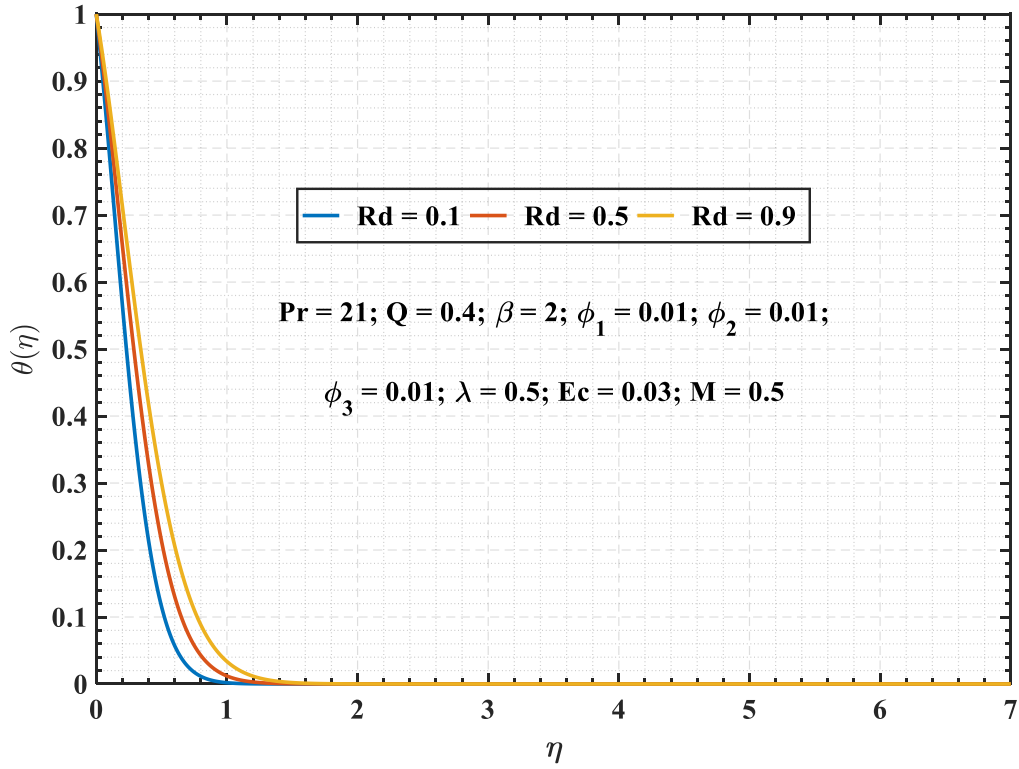

**Fig. 6.** $\theta(\eta)$ versus $Rd$.

As the heat source parameter $(Q)$ goes up, the temperature $\theta(\eta)$ of the C-M NF precipitously increases all over the thermal boundary layer. This phenomenon occurs because the magnitude of the $Q$ indicates a greater amount of heat generation within the fluid, leading to more energy being stored. The $\theta(\eta)$ of the nanofluid is amplified when the radiation parameter $(Rd)$ surges, which is owing to an increased radiative heat transfer effect. In the targeted drug release process, temperature sensitive nanoparticles release drugs as soon as local body parts reach elevated temperatures. This action enables the drug to become more efficient, especially during the hyperthermia-assisted treatment procedure. Normal heating ensures precisely distributed drugs to the affected areas, significantly improving treatment outcomes and minimally impacting healthy tissue.

The local skin friction coefficient $Cf_x Re_x^{1/2}$ is used as an indicator for calculating the interaction of nanofluid with the arterial wall in biomedical applications. More specifically, it directly measures the wall shear stress, which is proportional to velocity gradient on the surface. A higher value of $Cf_x Re_x^{1/2}$ reflects a larger amount of shear stress, which could lead to excessive mechanical force on the arterial walls, which may result in aggravated damage in stenosed areas. Conversely, a lower value signifies less shear stress, which could curtail particle wall interaction, leading to inadequate drug delivery. In targeted drug delivery, specifically through C-M NF modeled stenosed arteries, regulating skin friction ensures a balance, which guarantees enough interaction for effective drug transfer without placing harmful stress on the vessel wall. Thus, we can improve the accuracy and safety of therapeutic delivery systems based on nanoparticles by fine-tuning the flow characteristics by examining the maximum and minimum values of local skin friction. From Table 3, it is observed that the $Cf_x Re_x^{1/2}$ of C-M NF curtails with an increase in the volume fraction of nanoparticles. Per unit rise in $M$ and $\lambda$, the $Cf_x Re_x^{1/2}$ declines by 71.4% and 219%, respectively. $Cf_x Re_x^{1/2}$ of C-M NF enhances by 66.1% with a per unit rise in $\beta$, respectively.

**Table 3: Variation in $Cf_x Re_x^{1/2}$ for different parameter values.**

| $\lambda$ | $\beta$ | $M$ | $\phi_1$ | $\phi_2$ | $\phi_3$ | $Cf_x Re_x^{1/2}$ |
|---|---|---|---|---|---|---|
| 0.1 | 2 | 0.5 | 0.01 | 0.01 | 0.01 | -1.93244338 |
| 0.5 | 2 | 0.5 | 0.01 | 0.01 | 0.01 | -2.77675870 |
| 0.9 | 2 | 0.5 | 0.01 | 0.01 | 0.01 | -3.69025789 |
| **Slope of linear regression** | | | | | | **-2.19769491** |
| 0.5 | 1 | 0.5 | 0.01 | 0.01 | 0.01 | -3.20640020 |
| 0.5 | 2 | 0.5 | 0.01 | 0.01 | 0.01 | -2.77675870 |
| 0.5 | 3 | 0.5 | 0.01 | 0.01 | 0.01 | -2.61794437 |
| **Slope of linear regression** | | | | | | **0.66158934** |
| 0.5 | 2 | 0.1 | 0.01 | 0.01 | 0.01 | -2.47631952 |
| 0.5 | 2 | 0.5 | 0.01 | 0.01 | 0.01 | -2.77675870 |
| 0.5 | 2 | 0.9 | 0.01 | 0.01 | 0.01 | -3.04859459 |
| **Slope of linear regression** | | | | | | **-0.71445319** |
| 0.5 | 2 | 0.5 | 0.01 | 0.01 | 0.01 | -2.77675870 |
| 0.5 | 2 | 0.5 | 0.02 | 0.01 | 0.01 | -2.86352514 |
| 0.5 | 2 | 0.5 | 0.03 | 0.01 | 0.01 | -2.95266625 |
| **Slope of linear regression** | | | | | | **-8.70725512** |
| 0.5 | 2 | 0.5 | 0.01 | 0.01 | 0.01 | -2.77675870 |
| 0.5 | 2 | 0.5 | 0.01 | 0.02 | 0.01 | -2.95941672 |
| 0.5 | 2 | 0.5 | 0.01 | 0.03 | 0.01 | -3.14944049 |
| **Slope of linear regression** | | | | | | **-18.3557211** |
| 0.5 | 2 | 0.5 | 0.01 | 0.01 | 0.01 | -2.77675870 |
| 0.5 | 2 | 0.5 | 0.01 | 0.01 | 0.02 | -2.86447502 |
| 0.5 | 2 | 0.5 | 0.01 | 0.01 | 0.03 | -2.95459968 |
| **Slope of linear regression** | | | | | | **-8.80266946** |

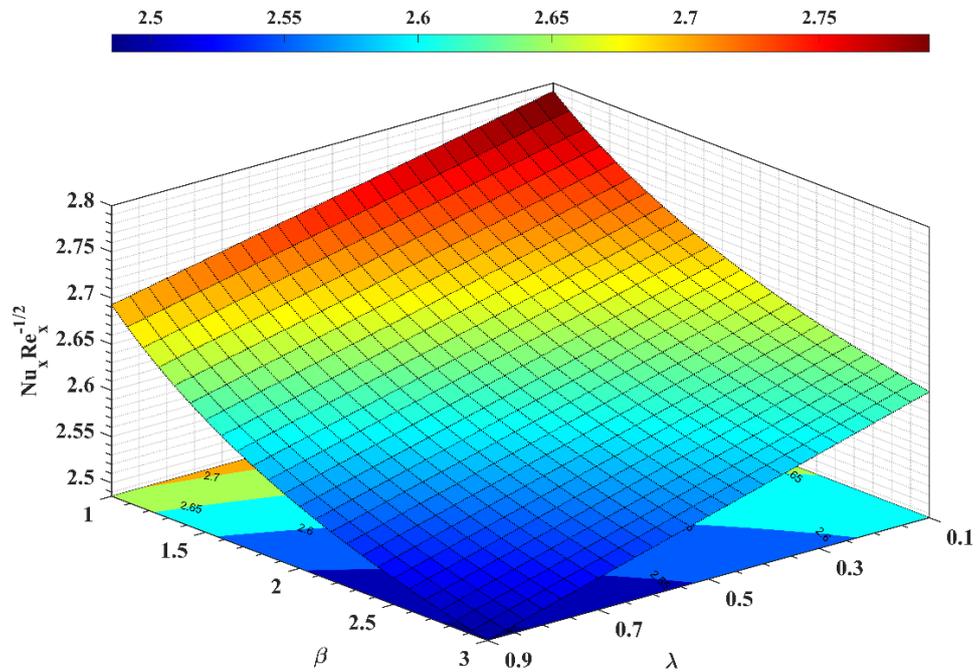

**Fig. 7.** Contour of $\beta$ and $\lambda$ on $Nu_x Re_x^{-1/2}$.

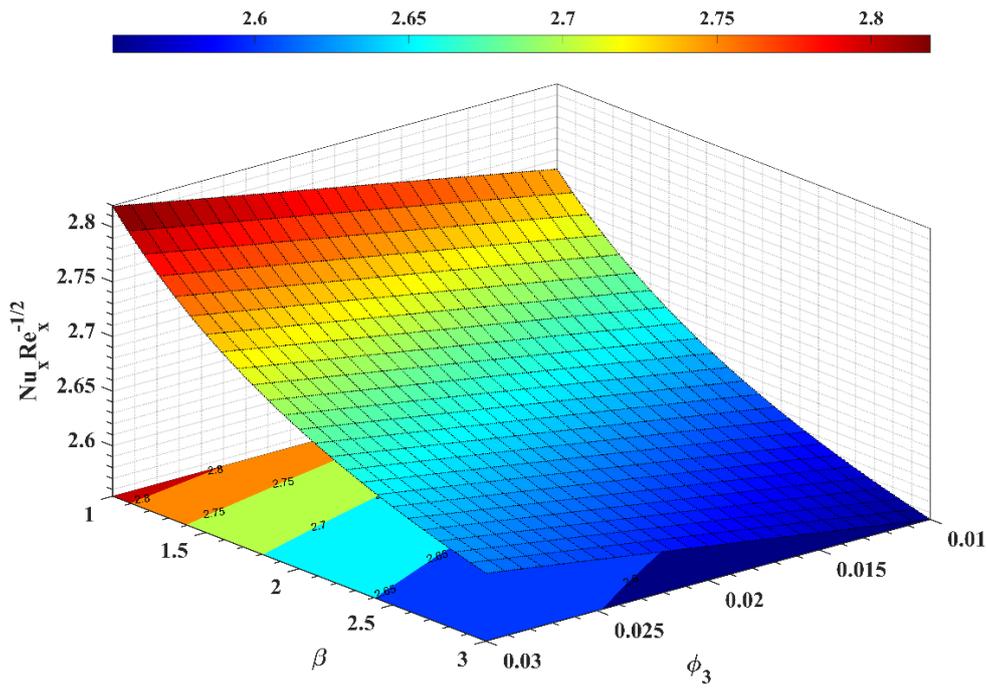

**Fig. 8.** Contour of $\beta$ and $\phi_3$ on $Nu_x Re_x^{-1/2}$.

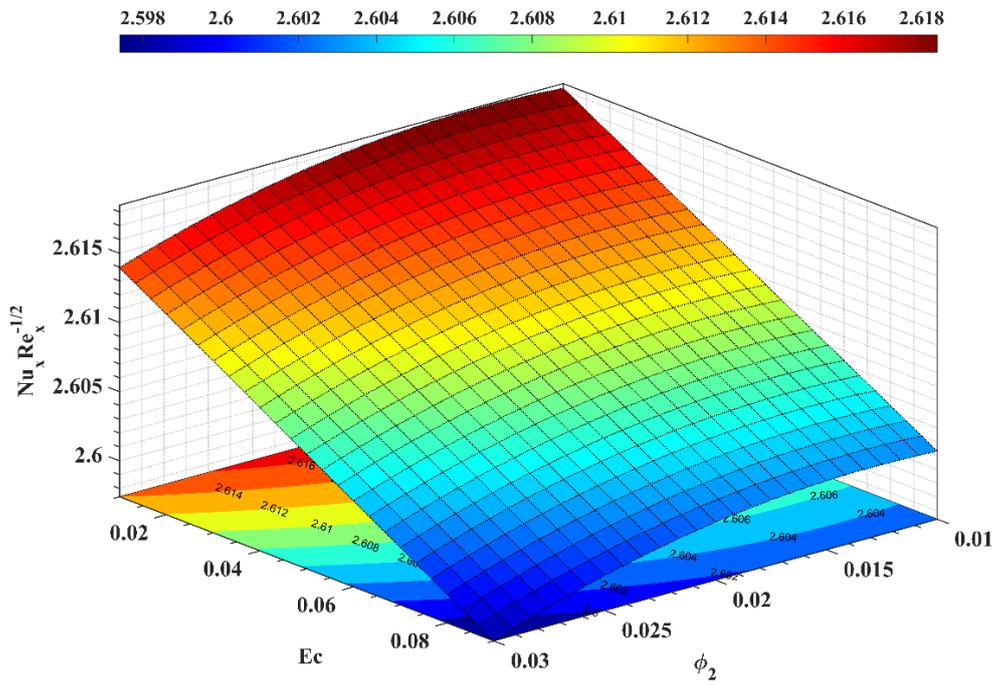

**Fig. 9.** Contour of $Ec$ and $\phi_3$ on $Nu_x Re_x^{-1/2}$.

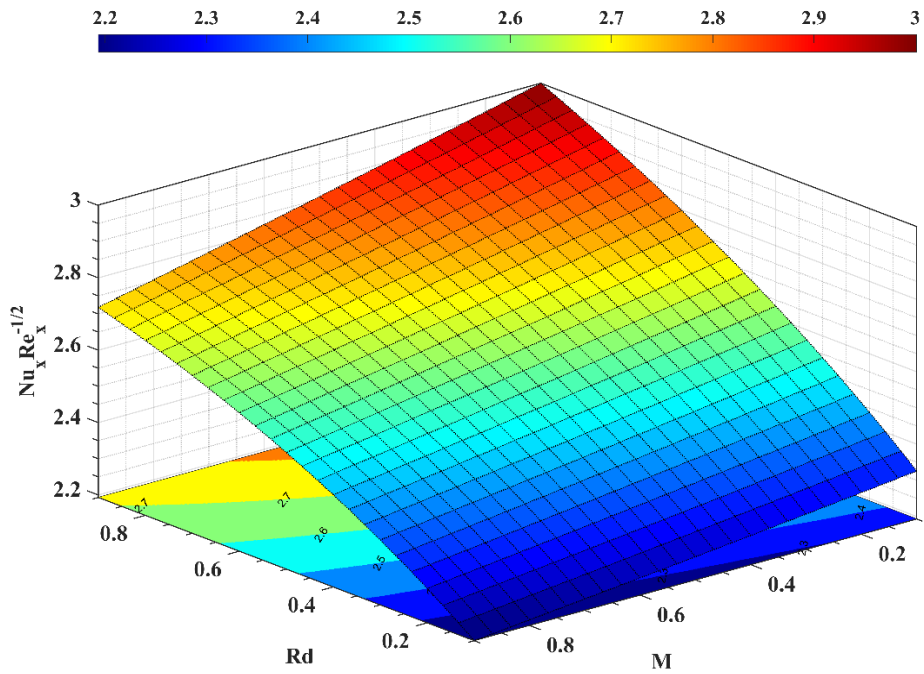

**Fig. 10.** Contour of $Rd$ and $M$ on $Nu_x Re_x^{-1/2}$.

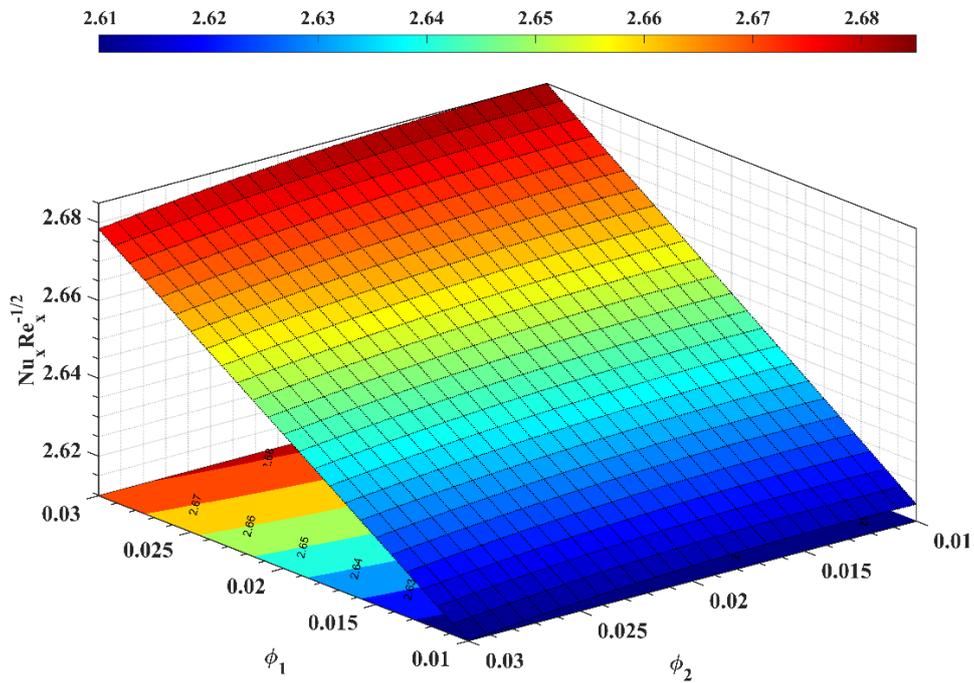

**Fig. 11.** Contour of $\phi_1$ and $\phi_2$ on $Nu_x Re_x^{-\frac{1}{2}}$.

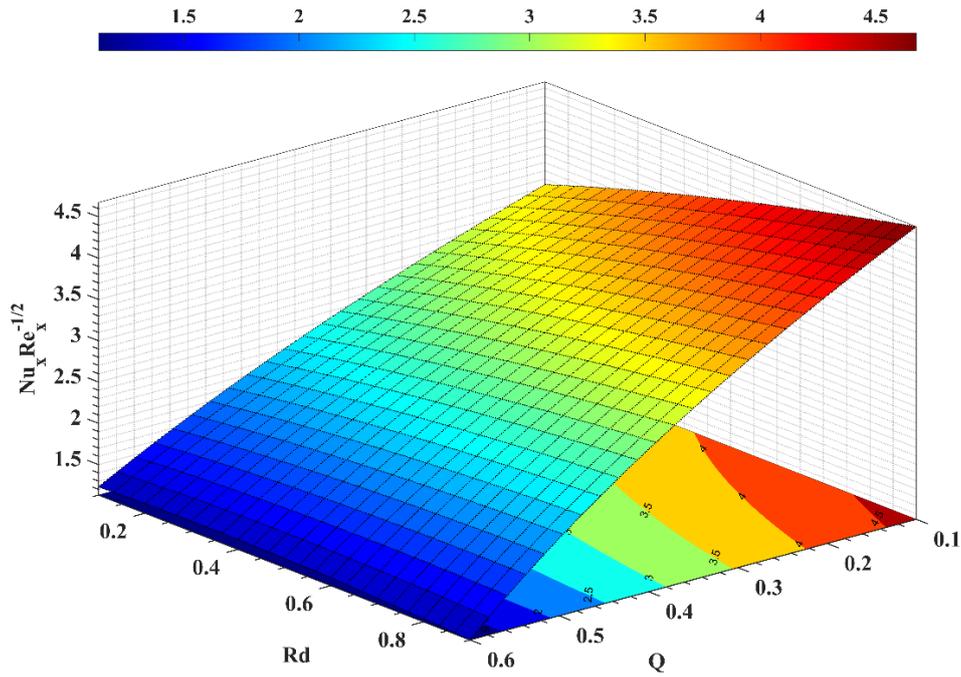

**Fig. 12.** Contour of $Rd$ and $Q$ on $Nu_x Re_x^{-1/2}$.

The local Nusselt number $Nu_x Re_x^{-1/2}$ is a dimensionless number that indicates convective heat transfer from arterial wall to the nanofluid and plays a crucial role in thermodynamics of life sciences. A higher $Nu_x Re_x^{-1/2}$ indicates better heat transfer, which is vital for activating the drug carriers that are thermally responsive and improving drug diffusion rates. However, excessive heat transfer can result in undesirable tissue heating if not regulated properly. Conversely, a lower $Nu_x Re_x^{-1/2}$ lowers heat exchange capacity, potentially impeding the temperature-sensitive mechanisms necessary for an effective drug release process. In drug delivery targeting applications, particularly within stenosed arteries, maintaining an optimal heat transfer rate is essential for thermal control. This facilitates the activation of drug release mechanisms at the correct location and time while minimizing thermal damage to healthy tissues. Therefore, to design efficient and safe heat assisted therapeutic systems, analyzing maximum and minimum values of $Nu_x Re_x^{-1/2}$ is crucial. The surface plots (Figs. 7-12) display the $Nu_x Re_x^{-1/2}$ under specific parameter combinations. It is observed that the $Nu_x Re_x^{-1/2}$ is high for the increasing parameter values of $Rd, \phi_1$, and $\phi_3$ and decreasing parameter values of $\beta, Q, Ec, M, \phi_2$, and $\lambda$.

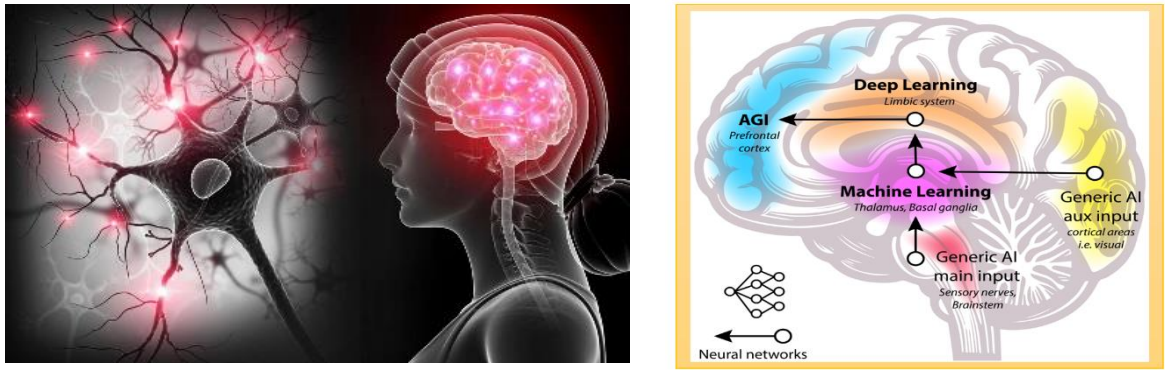

**Fig. 13** An artificial intelligence (AI) brain

## 5. Artificial Neural Network (ANN)

A network of linked computational nodes (neurons) that processes and transmits information, modeled after the architecture and functioning of the nervous system. Fig. 13 shows how important brain areas are similar to AI ideas, beginning with the primary input, a convergence of peripheral afferent inputs. A variety of neural pathways contribute to motion perception, including the red brain combination of nerves, the original sensitive nerves in the skull, and autonomic outputs from the nervous system that control emotions. The neuron, sometimes called a node, is a fundamental building block of computing. It takes in data, uses that data to determine its weights, and then uses a stimulation process to produce an output. ANN requires the following layers to create the final projection or result:

(i) **Layer 1 (Input):** Data from outside sources is received by this layer.

(ii) **Layer II (Invisible Layers):** Runs data through complex models

(iii) **Layer III (Output layer):** Production of the final prediction or result is the responsibility of the output layer.

To evaluate the efficacy of an artificial neural network model, we calculated its mean squared error (MSE) as follows.

$$\left| MSE = \frac{1}{N} \sum_{i=1}^{N} \left[ X_{Bvp4c} - X_{ANN(i)} \right]^2 \right. \tag{21}$$

here, $N$ is used to demonstrate that the total number of data points, $X_{bvp4c}$, denotes the real numerical results, and $X_{ANN}$ denotes predicted value produced by an ANN. The coefficient of determination is signified as R, and it is defined in Eq. 22, which examines the correctness of the model's approximations, and the error rate is calculated using Eq. 23.

$$\left| R = \sqrt{1 - \frac{\left[\sum_{i=1}^{N}\left[X_{Bvp4c} - X_{ANN(i)}\right]\right]^2}{\sum_{i=1}^{N}\left(X_{Bvp4c}\right)^2}} \right. \tag{22}$$

$$\text{Error rate}(\%) = \left( \frac{X_{Bvp4c} - X_{ANN}}{X_{Bvp4c}} \right) \times 100 \tag{23}$$

An ANN's design, including parameters, dataset size, and error distribution, significantly impacts the algorithm's precision and generalizability. To get the best results, ensure that the data set is appropriately scaled and processed. Additionally, fine-tuning important variables like the number of cells, learning rate, and activation functions is essential. For real-world applications like optimizing heat transfer or developing cooling techniques using tiny fluids, monitoring the error distribution is an excellent method to measure the framework's reliability and ensure it produces correct predictions. Employing a range of information sets, the efficacy of the ANN method was assessed, validated, and enhanced. 70% of the information was utilized for learning, 15% for examination, and 15% to evaluate the current model (see Fig. 14). The Levenberg-Marquardt artificial network performance on the Nusselt number for testing, validation, and training is demonstrated in Figs. 14–19. Fig.14 shows that the best validation occurs at the $4^{th}$ epoch, and the best validation performance is 0.19921. Very accurate predictions are shown by the fact that most of the errors are clustered around zero, as shown in Fig. 15. The biggest bar represents the training data (blue). From a biological perspective, this means that the algorithm accurately predicts the transfer of heat through convection, which is crucial for simulating the use of hyperthermia and other thermal treatments and the dispersion of drugs that are sensitive to heat. Fig. 16 shows that the performance on the gradient, Mu, validation checks, here clarified, represents that the predictable step-size at epoch 06 is around 1.443e-10. Fig. 17 shows that the model's overall performance has an R value of 0.99457, which is close to 1, implying that the current model produces accurate results. On the other hand, the convergence indicates that the algorithm learns to monitor and forecast continuous thermal activity. This is shown by its time series performance in training throughout epochs (see Fig. 18). One way to measure the Nusselt number in a biological context is to take thickness of boundary layer and plot it against time or space (see Fig. 19). For example, when a tumor area undergoes heat treatment, the Nusselt number is tracked. The autocorrelation error for the Nusselt number is presented in Fig. 20.

**Validation and Test Data**
Set aside some target timesteps for validation and testing.

**Select Percentages**

✦ Randomly divide up the 24 target timesteps:

| | | |
|---|---|---|
| 🟦 Training: | 70% | 16 target timesteps |
| 🟩 Validation: | 15% | 4 target timesteps |
| 🟥 Testing: | 15% | 4 target timesteps |

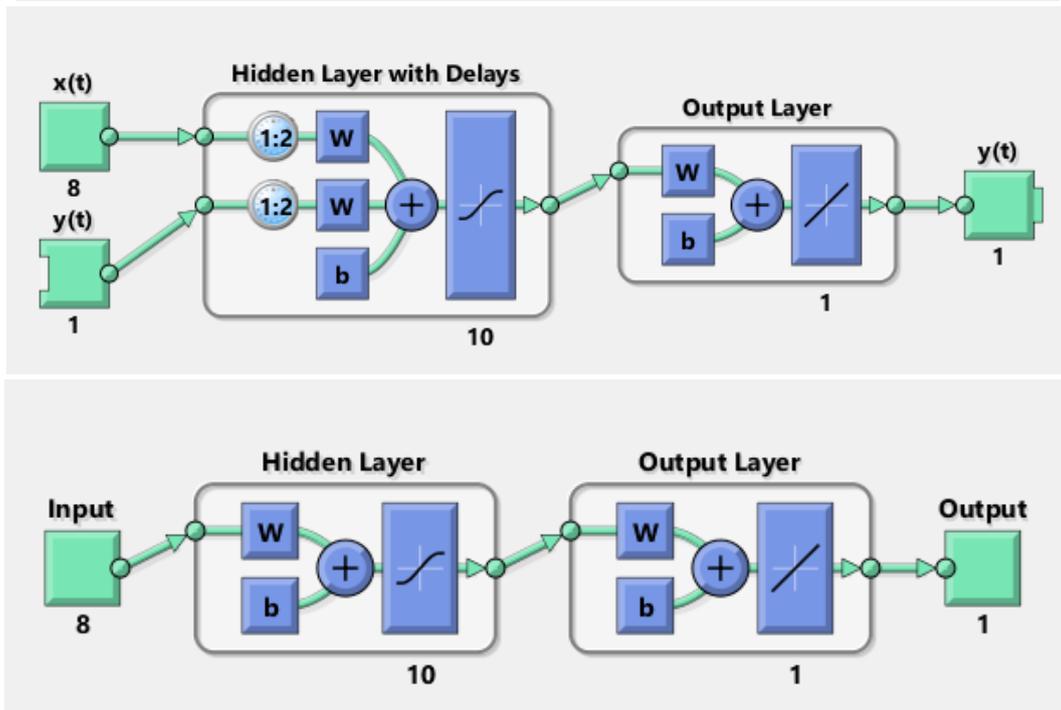

**Fig. 14.** Computer generated structure for ANN.

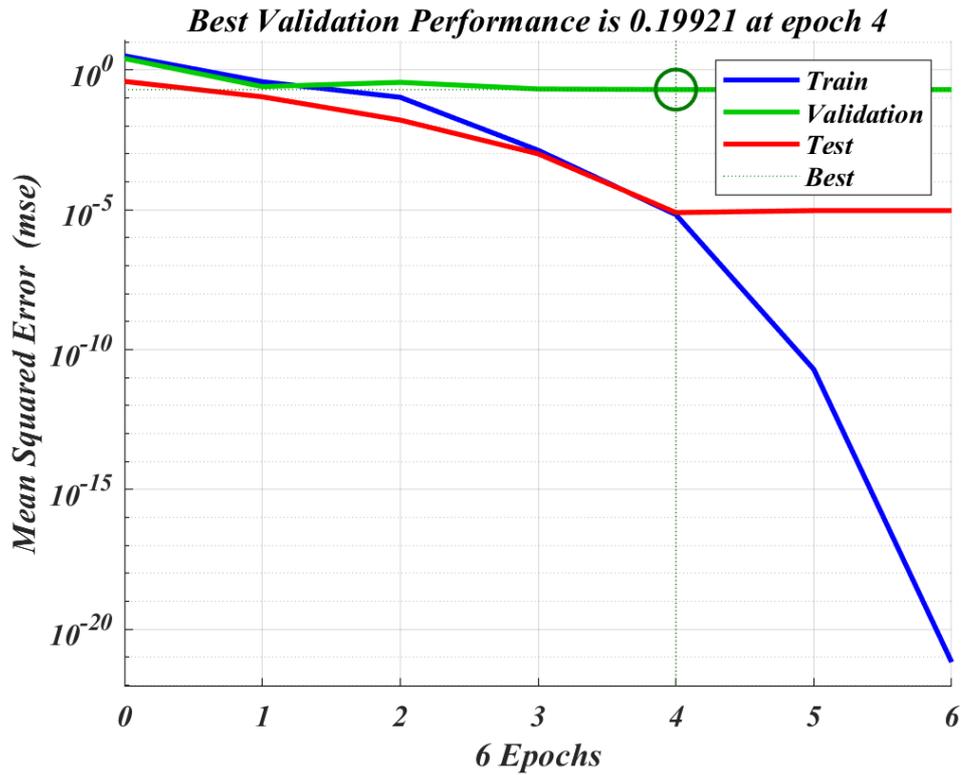

**Fig. 15**. Best validation Nusselt number.

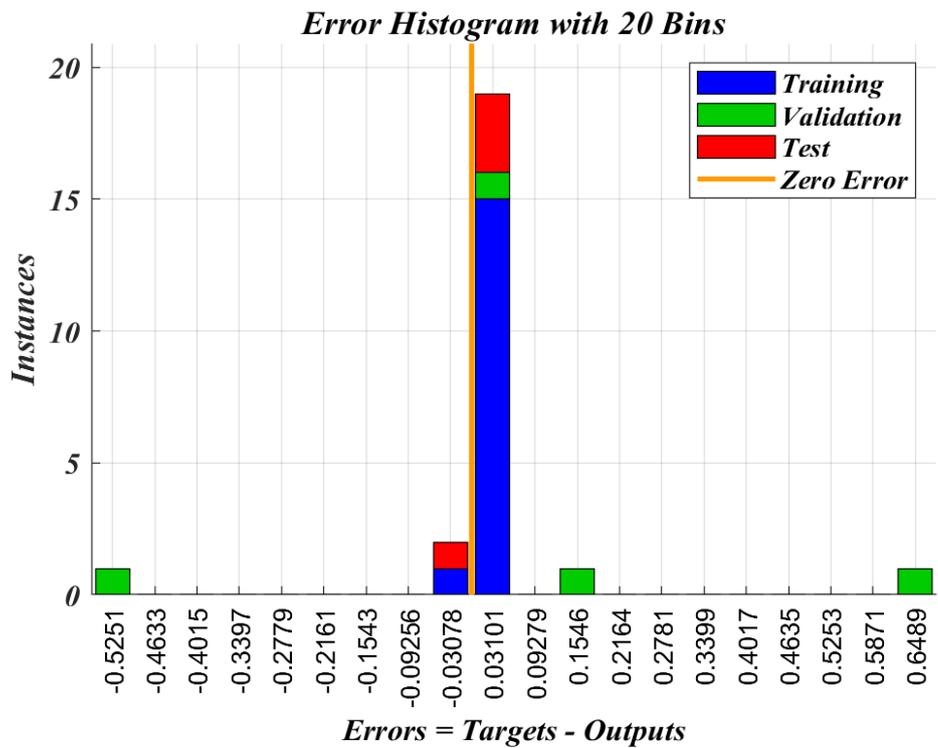

**Fig. 16.** Error Histogram for Nusselt number.

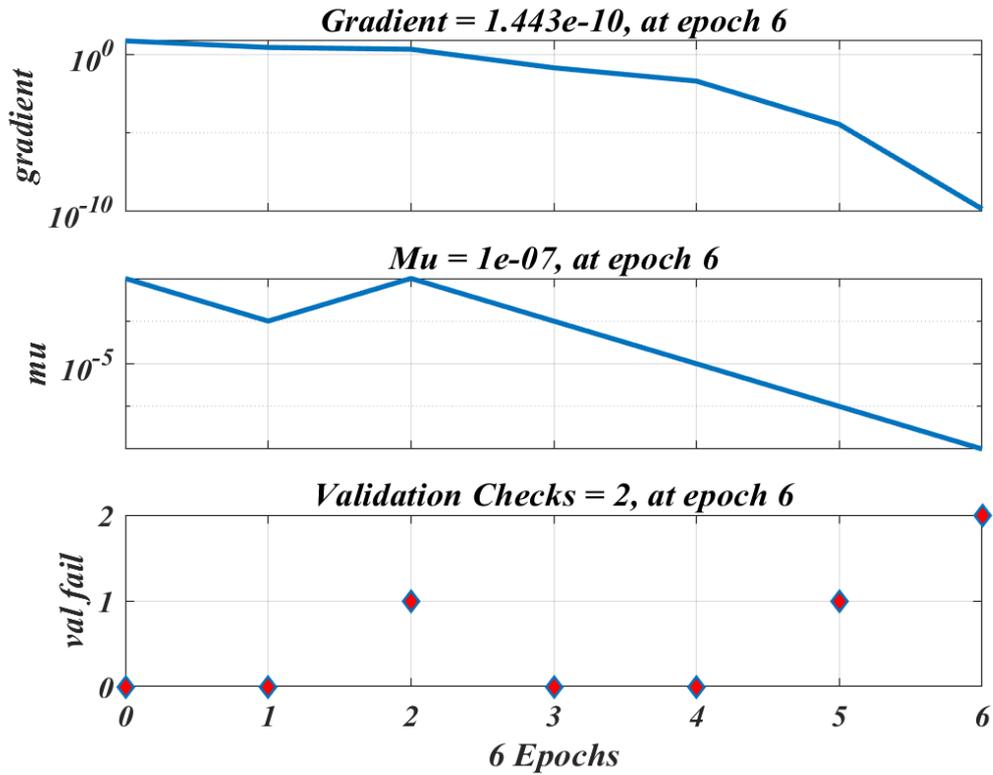

**Fig. 17.** Performance on the gradient, Mu, validation checks Nusselt number.

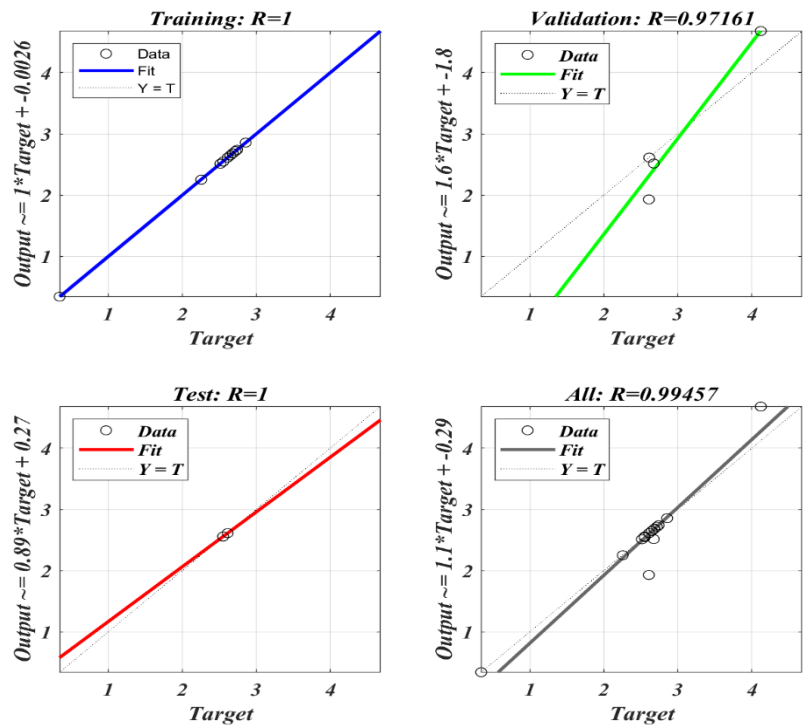

**Fig. 18.** Over all performance for Nusselt number.

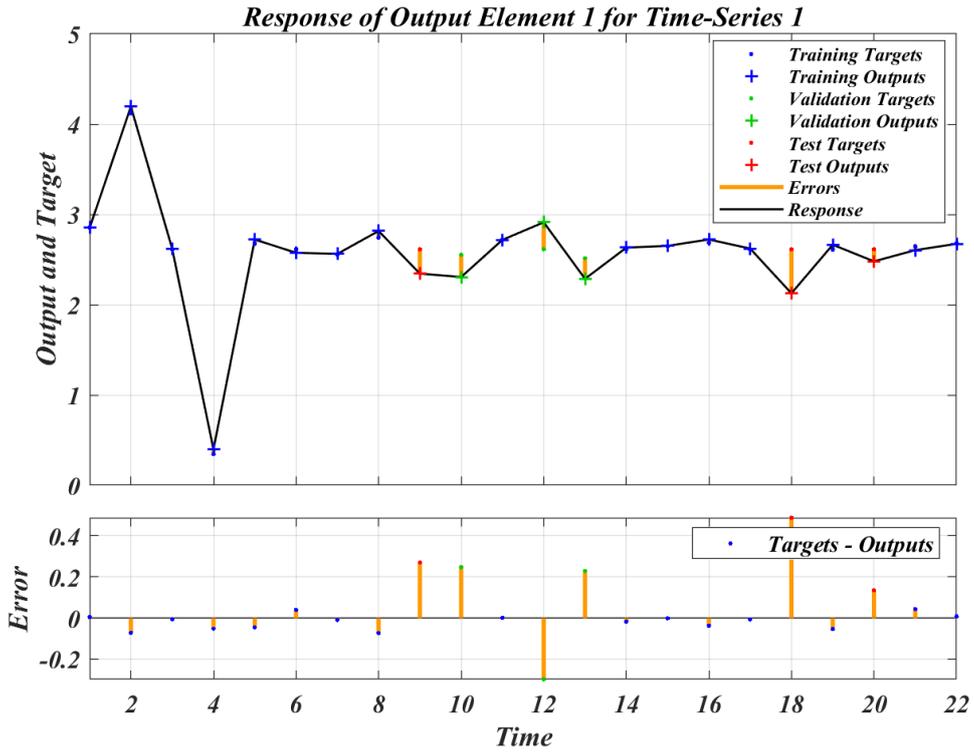

**Fig. 19.** Time series for Nusselt number.

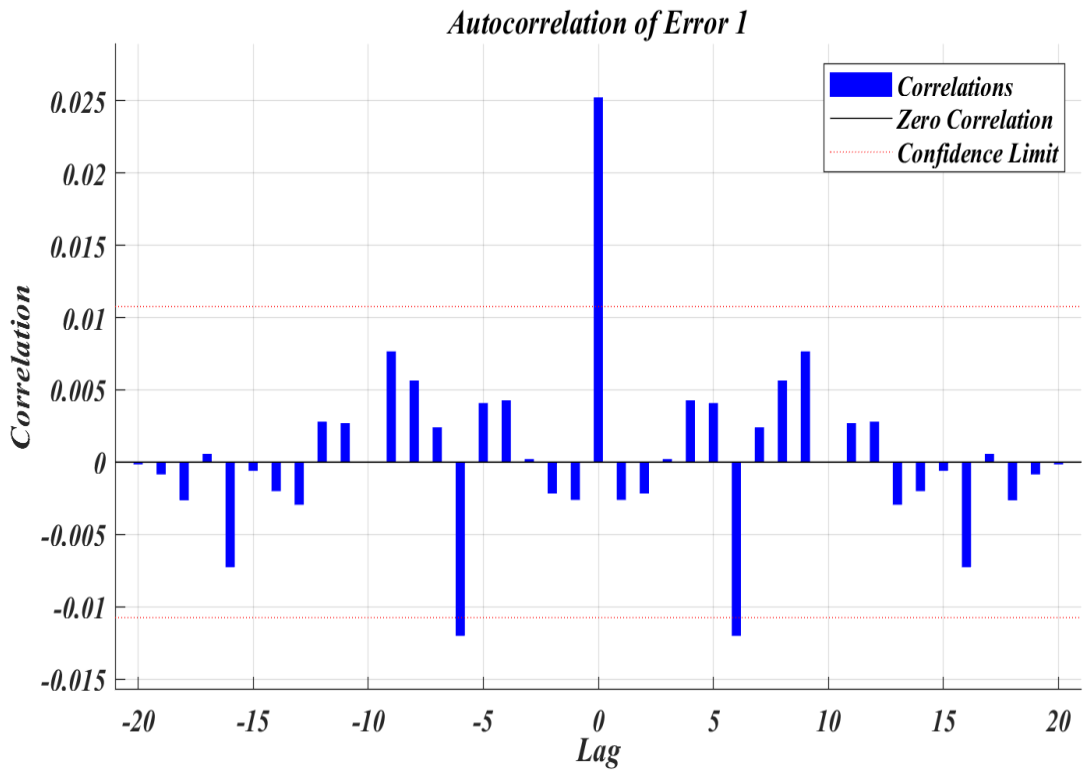

**Fig. 20.** Autocorrelation error for Nusselt number.

**Table 4: Comparison of $Nu_x Re_x^{-1/2}$ with different parameters.**

| Rd | Q | λ | β | M | $\phi_1$ | $\phi_2$ | $\phi_3$ | $Nu_x Re_x^{-1/2}$ BVP4C | ANN | Error |
|---|---|---|---|---|---|---|---|---|---|---|
| 0.1 | 0.4 | 0.5 | 2 | 0.5 | 0.01 | 0.01 | 0.01 | 2.25707588 | 2.28195112 | -0.024875245 |
| 0.5 | 0.4 | 0.5 | 2 | 0.5 | 0.01 | 0.01 | 0.01 | 2.61419013 | 2.61593053 | -0.001740394 |
| 0.9 | 0.4 | 0.5 | 2 | 0.5 | 0.01 | 0.01 | 0.01 | 2.85659094 | 2.87487292 | -0.018281983 |
| 0.5 | 0.1 | 0.5 | 2 | 0.5 | 0.01 | 0.01 | 0.01 | 4.12138101 | 3.67842785 | 0.442953158 |
| 0.5 | 0.4 | 0.5 | 2 | 0.5 | 0.01 | 0.01 | 0.01 | 2.61419013 | 2.61593053 | -0.001740394 |
| 0.5 | 0.7 | 0.5 | 2 | 0.5 | 0.01 | 0.01 | 0.01 | 0.34525708 | 0.35524389 | -0.009986806 |
| 0.5 | 0.4 | 0.1 | 2 | 0.5 | 0.01 | 0.01 | 0.01 | 2.67588873 | 2.68121837 | -0.005329639 |
| 0.5 | 0.4 | 0.5 | 2 | 0.5 | 0.01 | 0.01 | 0.01 | 2.61419013 | 2.61593053 | -0.001740394 |
| 0.5 | 0.4 | 0.9 | 2 | 0.5 | 0.01 | 0.01 | 0.01 | 2.55233738 | 2.49696504 | 0.055372334 |
| 0.5 | 0.4 | 0.5 | 1 | 0.5 | 0.01 | 0.01 | 0.01 | 2.74183725 | 2.78556657 | -0.043729325 |
| 0.5 | 0.4 | 0.5 | 2 | 0.5 | 0.01 | 0.01 | 0.01 | 2.61419013 | 2.61593053 | -0.001740394 |
| 0.5 | 0.4 | 0.5 | 3 | 0.5 | 0.01 | 0.01 | 0.01 | 2.55383002 | 2.51020564 | 0.043624378 |
| 0.5 | 0.4 | 0.5 | 2 | 0.1 | 0.01 | 0.01 | 0.01 | 2.71907485 | 2.71382887 | 0.005245983 |
| 0.5 | 0.4 | 0.5 | 2 | 0.5 | 0.01 | 0.01 | 0.01 | 2.61419013 | 2.61593053 | -0.001740394 |
| 0.5 | 0.4 | 0.5 | 2 | 0.9 | 0.01 | 0.01 | 0.01 | 2.51544257 | 2.52325891 | -0.007816342 |
| 0.5 | 0.4 | 0.5 | 2 | 0.5 | 0.01 | 0.01 | 0.01 | 2.61419013 | 2.61593053 | -0.001740394 |
| 0.5 | 0.4 | 0.5 | 2 | 0.5 | 0.02 | 0.01 | 0.01 | 2.64978191 | 2.64218148 | 0.007600428 |
| 0.5 | 0.4 | 0.5 | 2 | 0.5 | 0.03 | 0.01 | 0.01 | 2.68502888 | 2.68685869 | -0.001829803 |
| 0.5 | 0.4 | 0.5 | 2 | 0.5 | 0.01 | 0.01 | 0.01 | 2.61419013 | 2.61593053 | -0.001740394 |
| 0.5 | 0.4 | 0.5 | 2 | 0.5 | 0.01 | 0.02 | 0.01 | 2.61373367 | 2.61718477 | -0.003451103 |
| 0.5 | 0.4 | 0.5 | 2 | 0.5 | 0.01 | 0.03 | 0.01 | 2.60983068 | 2.47574753 | 0.134083158 |
| 0.5 | 0.4 | 0.5 | 2 | 0.5 | 0.01 | 0.01 | 0.01 | 2.61419013 | 2.61593053 | -0.001740394 |
| 0.5 | 0.4 | 0.5 | 2 | 0.5 | 0.01 | 0.01 | 0.02 | 2.64793655 | 2.70534501 | -0.057408462 |
| 0.5 | 0.4 | 0.5 | 2 | 0.5 | 0.01 | 0.01 | 0.03 | 2.68134734 | 2.6934915 | -0.012144165 |

## 6. Response surface methodology

Response Surface Methodology (RSM) is a robust statistical approach to explore and quantify the interactive effects of multiple pertinent parameters or independent variables on a specific response or dependent variable [46,47]. The data generated through Central Composite Design (CCD) is employed to construct and estimate the RSM model, enabling precise prediction and optimization of the targeted response. In this study, the drag coefficient of C-M NF is represented by the non-dimensional quantity $Cf_x Re_x^{1/2}$ is selected as the dependent variable (see Areekara et al. [48]). The Maxwell parameter ($0.1 \leq \lambda \leq 0.9$), Casson parameter ($1 \leq \beta \leq 3$), and Magnetic parameter ($0.1 \leq M \leq 0.9$) are treated as the independent variables influencing the $Cf_x Re_x^{1/2}$. Table 5 presents the key influencing parameters along with their corresponding levels. The response variable is modeled using a general equation that incorporates linear, quadratic, and interaction terms, which is given by:

$$Response = l_1 A + l_2 B + l_3 C + l_4 AB + l_5 BC + l_6 AC + l_7 A^2 + l_8 B^2 + l_9 C^2 + l_{10} \qquad (21)$$

The coefficients $l_i$ ($for\ i = 1\ to\ 10$) denote the regression parameters in the RSM model. Table 6 outlines experimental design and corresponding response values for 20 runs conducted based on CCD.

**Table 5:** Effective parameter levels

| Symbol | Parameter | Levels | | |
|---|---|---|---|---|
| | | -1 (Low) | 0 (Medium) | 1 (High) |
| A | $\lambda$ | 0.1 | 0.5 | 0.9 |
| B | $\beta$ | 1 | 2 | 3 |
| C | M | 0.1 | 0.5 | 0.9 |

**Table 6:** Experimental design with response

| Run | Coded Values | | | Actual Values | | | Response ($Cf_x Re_x^{1/2}$) | |
|---|---|---|---|---|---|---|---|---|
| | A | B | C | $\lambda$ | $\beta$ | M | Numerical value | Predicted value |
| 1 | -1 | -1 | -1 | 0.1 | 1 | 0.1 | -1.96020660 | -1.964457 |
| 2 | 1 | -1 | -1 | 0.9 | 1 | 0.1 | -3.84771604 | -3.848233 |
| 3 | -1 | 1 | -1 | 0.1 | 3 | 0.1 | -1.59998015 | -1.617657 |
| 4 | 1 | 1 | -1 | 0.9 | 3 | 0.1 | -3.14146018 | -3.129753 |
| 5 | -1 | -1 | 1 | 0.1 | 1 | 0.9 | -2.47378492 | -2.555593 |
| 6 | 1 | -1 | 1 | 0.9 | 1 | 0.9 | -4.63992016 | -4.692297 |
| 7 | -1 | 1 | 1 | 0.1 | 3 | 0.9 | -2.01980875 | -2.089273 |
| 8 | 1 | 1 | 1 | 0.9 | 3 | 0.9 | -3.78846900 | -3.854297 |
| 9 | -1 | 0 | 0 | 0.1 | 2 | 0.5 | -1.93244337 | -1.920845 |
| 10 | 1 | 0 | 0 | 0.9 | 2 | 0.5 | -3.69025789 | -3.745245 |
| 11 | 0 | -1 | 0 | 0.5 | 1 | 0.5 | -3.20640020 | -3.230025 |
| 12 | 0 | 1 | 0 | 0.5 | 3 | 0.5 | -2.61794437 | -2.637625 |
| 13 | 0 | 0 | -1 | 0.5 | 2 | 0.1 | -2.47631952 | -2.469005 |
| 14 | 0 | 0 | 1 | 0.5 | 2 | 0.9 | -3.04859459 | -3.126845 |
| 15 | 0 | 0 | 0 | 0.5 | 2 | 0.5 | -2.77675870 | -2.797925 |
| 16 | 0 | 0 | 0 | 0.5 | 2 | 0.5 | -2.77675870 | -2.797925 |
| 17 | 0 | 0 | 0 | 0.5 | 2 | 0.5 | -2.77675870 | -2.797925 |
| 18 | 0 | 0 | 0 | 0.5 | 2 | 0.5 | -2.77675870 | -2.797925 |
| 19 | 0 | 0 | 0 | 0.5 | 2 | 0.5 | -2.77675870 | -2.797925 |
| 20 | 0 | 0 | 0 | 0.5 | 2 | 0.5 | -2.77675870 | -2.797925 |

**Table 7:** ANOVA table.

|  | Degree of freedom | Adjusted sum of squares | Adjusted mean squares | F-Value | p-Value |
|---|---|---|---|---|---|
| **Model** | 9 | 10.2862 | 1.14291 | 3485.90 | 0.000 |
| *Linear* | 3 | 10.0640 | 3.35466 | 10231.75 | 0.000 |
| $\lambda$ | 1 | 8.3204 | 8.32036 | 25377.20 | 0.000 |
| $\beta$ | 1 | 0.8764 | 0.87638 | 2672.96 | 0.000 |
| $M$ | 1 | 0.8672 | 0.86724 | 2645.09 | 0.000 |
| *Square* | 3 | 0.1140 | 0.03801 | 115.93 | 0.000 |
| $\lambda * \lambda$ | 1 | 0.0034 | 0.00339 | 10.35 | 0.009 |
| $\beta * \beta$ | 1 | 0.0508 | 0.05082 | 155.01 | 0.000 |
| $M * M$ | 1 | 0.0005 | 0.00052 | 1.59 | **0.236** |
| *2-Way Interaction* | 3 | 0.1082 | 0.03607 | 110.02 | 0.000 |
| $\lambda * \beta$ | 1 | 0.0691 | 0.06910 | 210.76 | 0.000 |
| $\lambda * M$ | 1 | 0.0320 | 0.03198 | 97.54 | 0.000 |
| $\beta * M$ | 1 | 0.0071 | 0.00714 | 21.77 | 0.001 |
| Error | 10 | 0.0033 | 0.00033 |  |  |
| Lack-of-Fit | 5 | 0.0033 | 0.00066 | * | * |
| Pure Error | 5 | 0.0000 | 0.00000 |  |  |
| Total | 19 | 10.2895 |  |  |  |
|  | $R^2 = 99.97\%$ |  |  | Adjusted $R^2 = 99.94\%$ |  |

The ANOVA results are presented in Tables 7, which assess the effectiveness of the fitted RSM model. Statistical significance is established for a parameter if its p-value falls below the threshold of 0.05 and its F-value exceeds 1 (see Areekara et al. [49]). Since the quadratic term $M^2$ does not exhibit statistical significance, hence it is excluded from the model. The model demonstrates a high accuracy, as it is indicated by the coefficient of determination ($R^2$), which is found to be 99.94%.

The fitted quadratic models for $Cf_x Re_x^{1/2}$ in uncoded form are given by:

$$Cf_x Re_x^{1/2} = -2.2292 - 2.3280\,\lambda + 0.6863\,\beta - 0.7741\,M - 0.2195\,\lambda^2 - 0.1359\,\beta^2 + 0.0861\,M^2 + 0.2323\,\lambda * \beta - 0.3952\,\lambda * M + 0.0747\,\beta * M \tag{23}$$

Fig. 20 shows the residual versus observation order plot, which is used to further assess the estimated model's reliability. Additionally, the residual analysis reveals a maximum error of 0.04 in the fitted versus residual plot, enhancing the model's accuracy. By setting the third parameter at the mid-level, Figs. 21-23 (surface plots) and 24-26 (contour plots) demonstrate how two independent variables simultaneously impact the response variable. It can be seen from Figs. 21-26 that smaller values of $M$, $\lambda$, and larger values of $\beta$ result in the high $Cf_x Re_x^{1/2}$.

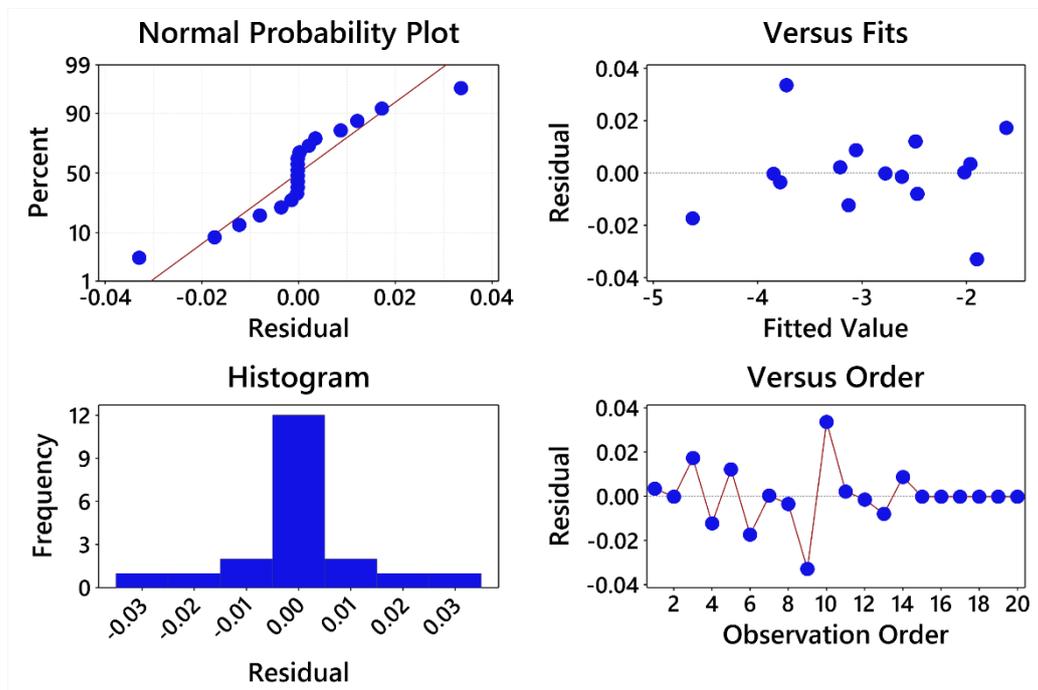

**Fig. 21.** Residual versus observation order plot of $Cf_x Re_x^{1/2}$.

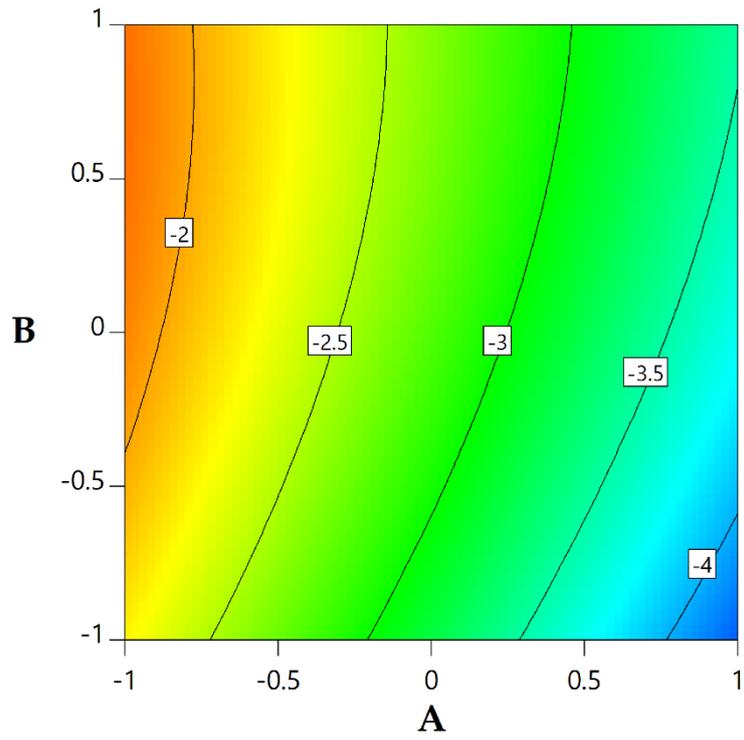

**Fig. 22.** Surface plots for $Cf_x Re_x^{1/2}$.

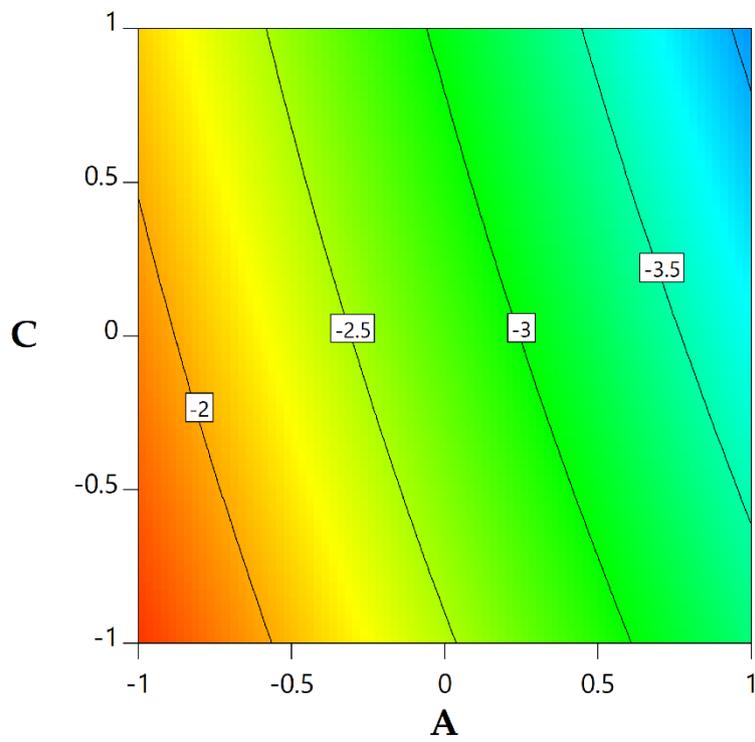

**Fig. 23.** Surface plots for $Cf_x Re_x^{1/2}$.

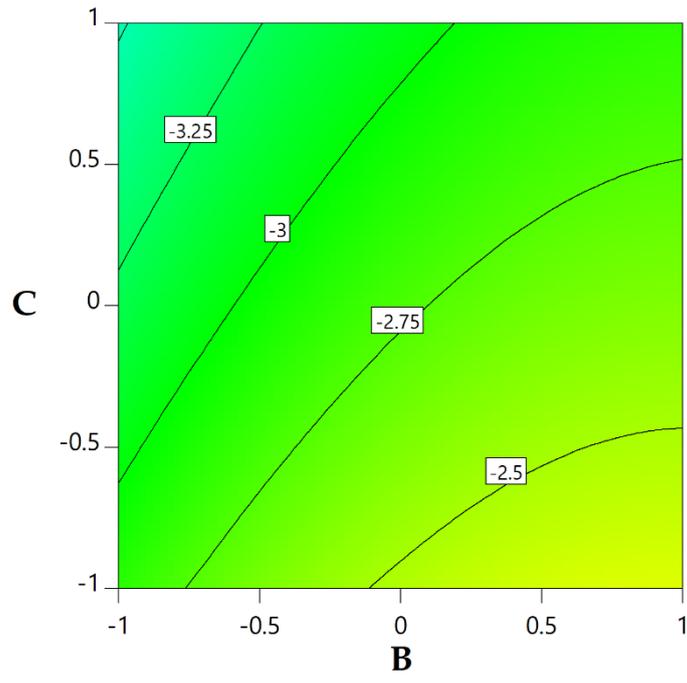

**Fig. 24.** Surface plots for $Cf_x Re_x^{1/2}$.

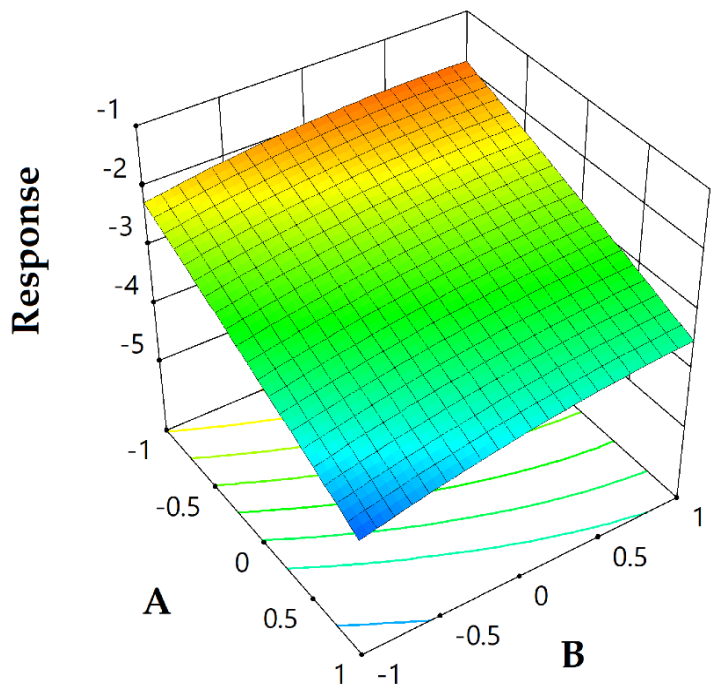

**Fig. 25.** Contour plots for $Cf_x Re_x^{1/2}$.

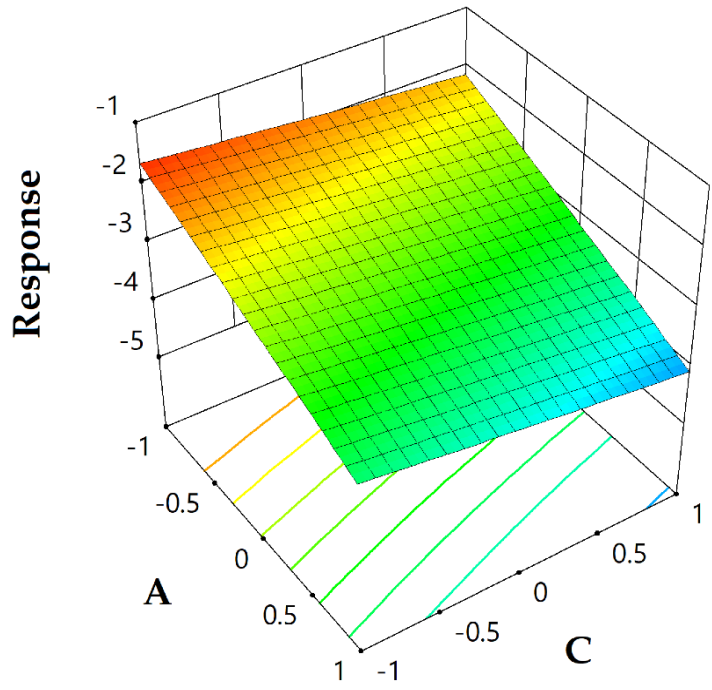

**Fig. 26.** Contour plots for $Cf_x Re_x^{1/2}$.

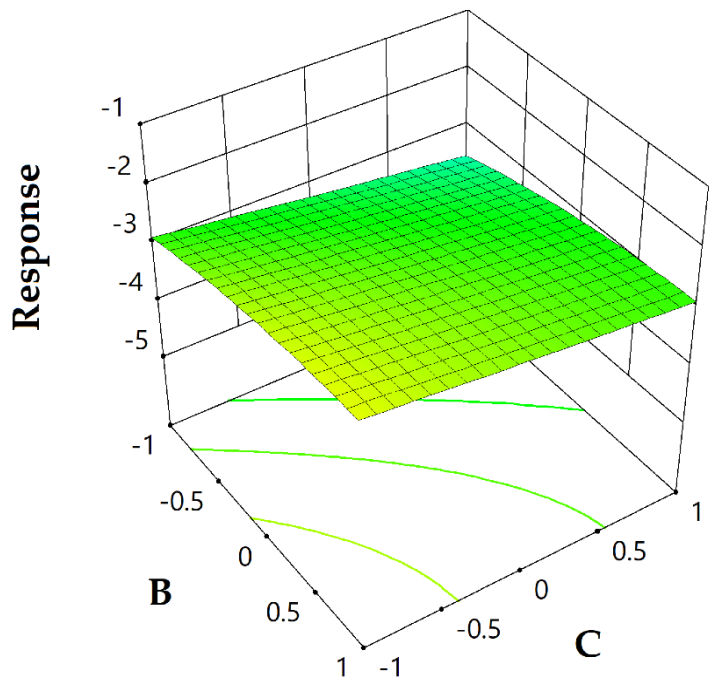

**Fig. 27.** Contour plots for $Cf_x Re_x^{1/2}$.

**Table 8:** Sensitivity of response $Cf_x Re_x^{1/2}$.

| A | B | C | Sensitivity Functions | | |
|---|---|---|---|---|---|
| | | | **Towards A** | **Towards B** | **Towards C** |
| -1 | -1 | -1 | -0.87163 | 0.445117 | -0.261131 |
| | | 0 | -0.934856 | 0.474985 | -0.261131 |
| | | 1 | -0.998082 | 0.504853 | -0.261131 |
| | 0 | -1 | -0.778692 | 0.173231 | -0.231263 |
| | | 0 | -0.841918 | 0.203099 | -0.231263 |
| | | 1 | -0.905144 | 0.232967 | -0.231263 |
| | 1 | -1 | -0.685754 | -0.098655 | -0.201395 |
| | | 0 | -0.74898 | -0.068787 | -0.201395 |
| | | 1 | -0.812206 | -0.038919 | -0.201395 |
| 0 | -1 | -1 | -0.941872 | 0.538055 | -0.324357 |
| | | 0 | -1.005098 | 0.567923 | -0.324357 |
| | | 1 | -1.068324 | 0.597791 | -0.324357 |
| | 0 | -1 | -0.848934 | 0.266169 | -0.294489 |
| | | 0 | -0.91216 | 0.296037 | -0.294489 |
| | | 1 | -0.975386 | 0.325905 | -0.294489 |
| | 1 | -1 | -0.755996 | -0.005717 | -0.264621 |
| | | 0 | -0.819222 | 0.024151 | -0.264621 |
| | | 1 | -0.882448 | 0.054019 | -0.264621 |
| 1 | -1 | -1 | -1.012114 | 0.630993 | -0.387583 |
| | | 0 | -1.07534 | 0.660861 | -0.387583 |
| | | 1 | -1.138566 | 0.690729 | -0.387583 |
| | 0 | -1 | -0.919176 | 0.359107 | -0.357715 |
| | | 0 | -0.982402 | 0.388975 | -0.357715 |
| | | 1 | -1.045628 | 0.418843 | -0.357715 |
| | 1 | -1 | -0.826238 | 0.087221 | -0.327847 |
| | | 0 | -0.889464 | 0.117089 | -0.327847 |
| | | 1 | -0.95269 | 0.146957 | -0.327847 |

## 7. Sensitivity analysis

Sensitivity analysis is a statistical method used to examine the magnitude and nature of dependency excreted by the independent variables on the response variable (see Fayyadh et al. [50]). After removing the insignificant terms, the quadratic model for $Cf_x Re_x^{1/2}$ in uncoded form is as follows:

$$Cf_x Re_x^{1/2} = -2.77655 - 0.91216\,A + 0.296037\,B - 0.294489\,C - 0.035121\,A^2 - 0.135943\,B^2 + 0.092938\,AB + 0.029868\,BC - 0.063226\,AC \tag{24}$$

The sensitivity functions of $Cf_x Re_x^{1/2}$ are:

With respect to A $= -0.91216 + 0.092938\,B - 0.063226\,C - 2*0.035121\,A$ \hfill (25)

With respect to B $= 0.296037 + 0.092938\,A + 0.029868\,C - 2*0.135943\,B$ \hfill (26)

With respect to C $= -0.294489 - 0.063226\,A + 0.029868\,B$ \hfill (27)

The sensitivities of $Cf_x Re_x^{1/2}$ are presented in Table 8. It is observed that both $\lambda$ and $M$ exhibit negative influence towards $Cf_x Re_x^{1/2}$. Natably the drag coefficient is most sensitive to the variations in $\lambda$.

## 8. Conclusion

The current investigation examines the CM NF through a stenosed artery modelled as a stretching surface while considering the effects of a thermal radiation, magnetic field, and a linear heat source. The findings contribute to developing advanced, non-invasive treatment strategies in biomedical engineering. Additionally, this research aligns with several United Nations Sustainable Development Goals, particularly SDG 3 SDG 9 by implementing, improving medical treatments, and SDG 9, implementing the latest and most innovative fluid modeling techniques. Therefore, besides providing technical insights into nanofluid-based drug delivery, the study supports global efforts toward sustainable and innovative healthcare solutions. Lastly, to predict the optimal heat transfer rate, machine learning artificial neural networks were utilized. This study reveals the following conclusions:

- The CM NF has a lower velocity than the Casson fluid due to the integrated effects of viscoelasticity and non-Newtonian characteristics.
- The velocity of CM NF is minimized by a rise in CM and magnetic parameters.
- The fluid's temperature is amplified with the rise in the heat source and radiation parameters.
- An enhanced heat transfer rate is observed with increasing volume fractions of copper and aluminum oxide nanoparticles. In contrast, a higher concentration of silver nanoparticles leads to a curtailment in heat transfer efficiency.
- The drag coefficient declines by 219% and increases by 66.1% for a unit rise in the Maxwell and Casson parameters, respectively.
- A better predictive outcome is shown by the machine learning method used for heat transfer rate with assistance of multiple factors.
- The overall R-value of the trained model was found to be 0.99457, which implies the results are significant for the suggested range of the contributing factors.
- The drag coefficient is most sensitive to the changes in the Maxwell parameter.

**DECLARATIONS**

**Data Availability Statement:** "The authors confirm that the data supporting the findings of this study are available within the article".

**Ethical Approval:** "Not applicable"

**Funding:** "Not applicable"